\documentclass[pra, notitlepage, twocolumn, superscriptaddress]{revtex4-1}

\usepackage{amsmath}
\usepackage{amsfonts}
\usepackage{physics}
\usepackage{xcolor}
\usepackage[T1]{fontenc}
\usepackage{mathptmx}
\usepackage{dcolumn}
\usepackage{eucal}
\usepackage{bm,bbm}
\usepackage{graphicx}
\usepackage{dsfont}
\usepackage{suffix}
\usepackage{hyperref}

\newcommand{\acom}[2]{\left\{#1,#2\right\}}

\newcommand{\ucwf}{\tilde\phi\up}
\newcommand{\cwf}{\phi\up}
\newcommand{\ui}{^{(i)}}

\newcommand{\proj}[2]{\Tr[\hat{\Pi}(#2) \, #1]}

\newcommand{\projE}[2]{\Tr_\text{E}[\hat{\Pi}^\text{E}(#2) \, #1]}
\newcommand{\ucdm}{\hat{\tilde\rho}\up}

\newcommand{\HS}{\hat{H}_{\rm S}}

\newcommand{\HE}{\hat{H}_{\rm E}}
\newcommand{\HI}{\hat{H}_{\text{int}}}
\newcommand{\evv}[1]{\left\langle\left\langle #1 \right\rangle\right\rangle}
\WithSuffix\newcommand\evv*[1]{\langle\langle #1 \rangle\rangle}
\WithSuffix\newcommand\Tr*[1]{\mathrm{Tr}\{ #1 \}}

\newcommand{\up}{^{(\Psi)}}

\newcommand{\sumss}{\sum_{s\in\{\uparrow,\downarrow\}}\sum_{s'\in\{\uparrow,\downarrow\}}}
\newcommand{\evy}[1]{\ev*{#1}{\vec{Y}_t}_\text{E}}
\newcommand{\evys}[1]{\ev*{#1}{\vec{Y}_t^{(\sigma)}}_\text{E}}
\newcommand{\us}{^{(\sigma)}}

\begin{document}

\title{{Contributions to single-shot energy exchanges in open quantum systems}}

\author{R. Sampaio}
\email{rui.ferreirasampaio@aalto.fi}
\affiliation{QTF Center of Excellence, Department of Applied Physics, Aalto University, P.O. Box 11000, FI-00076 Aalto, Finland}
\author{J. Anders}
\author{T. G. Philbin}
\affiliation{CEMPS, Physics and Astronomy, University of Exeter, Exeter, EX4 4QL, United Kingdom.}
\author{T. Ala-Nissila}
\affiliation{QTF Center of Excellence, Department of Applied Physics, Aalto University, P.O. Box 11000, FI-00076 Aalto, Finland}
\affiliation{Interdisciplinary Centre for Mathematical Modelling and Department of Mathematical Sciences, Loughborough University, Loughborough, Leicestershire LE11 3TU, UK}

\begin{abstract}
	The exchange of energy between a classical open system and its environment
	can be analysed for a single run of an experiment using the phase space trajectory of the system.
	By contrast, in the quantum regime such energy exchange processes must be defined for an {\it ensemble} of runs of the same experiment
	based on the reduced system density matrix.
	Single-shot approaches have been proposed for quantum systems that are weakly coupled to a heat bath.
	However, a single-shot analysis for a quantum system that is entangled or strongly interacting with external degrees
	of freedom has not been attempted because no system wave function exists for such a system within the standard formulation of quantum theory.
	Using the notion of the {\it conditional} wave function of a quantum system, we derive here an exact formula for the
	rate of total energy change in an open quantum system, valid for arbitrary coupling between the system and the environment. In particular,
	this allows us to identify three distinct contributions to the total energy flow:
	an external contribution coming from the explicit time dependence of the Hamiltonian,
	an interaction contribution associated with the interaction part of the Hamiltonian,
	and an entanglement contribution, directly related to the presence of entanglement between the system and its environment.
	Given the close connection between weak values and the conditional wave function, the
	approach presented here provides a new avenue for experimental studies of energy fluctuations in open quantum systems.
\end{abstract}

\date{\today}

\maketitle

\section{Introduction}

Open quantum systems are ubiquitous in realistic physical scenarios such as novel quantum devices and quantum computation, and a proper understanding of their behaviour is of both conceptual and practical importance. The focus of open quantum systems is to describe the non-unitary dynamics of a {system} embedded in a larger environment. In principle this can be done through
the system's reduced density operator, which contains the information required to compute the statistics of any observable of the system.
However, in dealing with large environments such as an infinite heat bath, the exact evolution of the reduced density operator is not available
and approximations must be employed \cite{oqs}. For example, the popular Lindblad master equation for
Markovian dynamics~\cite{Lindblad1983} can be derived under the assumption of weak coupling of the system to the
environment and a clear separation of time scales~\cite{oqs,rivas12}. Energy exchange
between the system and its environment can then be studied at the level of the system's reduced density operator~\cite{Vinjanampathy16}.

Recently, there has been growing interest in understanding the fluctuation of energy in open systems at the single-shot level of \textit{individual} runs of an
experimental protocol~\cite{Hekking:2013aa,Horowitz2012,Elouard17,Pekola2013,Pekola2016,sampaio16,Vinjanampathy16}.
This scenario is commonly described by unraveling master equations, leading to a dynamical equation for the evolution of a stochastic wave function~\cite{oqs,Gardiner92}. This stochastic state is used  as a numerical tool to calculate the density operator but it contains information
beyond it. In particular, approaches {based on the} quantum-jump method~\cite{Molmer93} have facilitated the definition of
stochastic thermodynamic quantities in an effort to extend the framework of classical stochastic thermodynamics to the quantum regime.
The explicit time dependence of the Hamiltonian is typically understood as the {external} work while jumps are associated with some form
of heat or entropy exchange due to the interaction with the
environment~\cite{Hekking:2013aa, Horowitz2012, Elouard17, Pekola2016, manzano17, Manikandan18}.
However, the quantum jump approach is based on the weak coupling limit which pre-empts the study of e.g. entanglement at the level of individual runs of an experiment \footnote{At the level of the reduced density operator the role of entanglement is actively studied, see \textit{e.g.}, \cite{Vinjanampathy16}.}.

In this paper our aim is to go beyond the weak coupling approximation to fully uncover the single-shot contributions to the energy
exchange of a quantum system that arise due to its
interaction and/or entanglement with its environment.
It should be noted, however, that tackling this question within the standard formulation of quantum mechanics is challenging
because a system that is entangled with its environment cannot be described by a wave function of its own, i.e., there is no system wave function that depends only on the degrees of freedom of the system \cite{conceptual18}. Within this formulation, the most detailed
description of a system is in terms of its reduced density matrix which is in general a mixture of wave functions.
A wave function for an open quantum system is here only admissible in the limit of a continuously measured
environment and Markovian evolution of the system \cite{Gambetta03, Gambetta02}.

In contrast it has been shown that a unique {\it conditional wave function} (CWF)~\cite{teufel2009bohmian} can be identified for a
quantum system that is entangled with its environment within the Bohmian phase space formulation of quantum mechanics~\cite{holland95,Bohm:1952aa,Bohm:1952ab,teufel2009bohmian}.
Conditional wave functions have become a tool for the investigation of transport in nanoscale electronic system~\cite{Oriols17,Abedi18,Oriols07,Zhan18}, chemical reactions~\cite{Benseny2014Applied} and the description of experiments such as the double slit~\cite{Mahler2016,Kocsis2011}, non-local steering~\cite{Xiao2017} and spin measurements~\cite{Hiley_2018}.

Here we employ the conditional wave function and its time evolution to define a {\it conditional energy} associated with a single-shot
experiment and establish a formally exact analytic expression for energy exchanges during the joint system-environment time evolution.
We show that {the rate of energy exchange} naturally
partitions into {three} terms that can be interpreted as an external, an interaction and an entanglement contribution.
Each of these terms can be non-zero while the others vanish, e.g., if the system and environment are entangled but there is no interaction
term between them, the entanglement contribution is non-zero while the interaction contribution is zero. By explicitly solving a few
simple model systems we illustrate the behaviour of these terms for various scenarios of driven open systems.
In contrast to many previous studies, the results presented here are valid for arbitrary environments -- not just heat baths --
and general Hamiltonians, including time-dependent interactions. In other words, no assumptions or approximations need to be made
about the environment or the
system's coupling to it, and also no coarse-graining (partial tracing) is needed.
The results provide a direct link between entanglement and energy fluctuations at the level of individual runs of an experiment. Given the close relationship between CWFs and weak measurements~\cite{norsen14} this provides a new avenue for empirical inquiry into the role of entanglement in energy fluctuations in open systems.

The paper is organized as follows. In section \ref{sec:cwf} we review the definition of the conditional wave function and its dynamics under a generalised Schr\"odinger equation.
In section \ref{sec:cond_energy} we formally define the conditional energy and show how its time derivative leads to three different contributions to the energy exchange, with different physical interpretations.
In section \ref{sec:examples} we analytically solve a number of examples to illustrate how these different contributions manifest themselves in concrete settings.
{We provide a generalization to mixed states in the Appendix and summarise our findings in Sec. \ref{sec:conclusions}.

\section{Conditional wave functions}\label{sec:cwf}

Here we give the definition and some intuition for the concept of CWF. A detailed discussion on the CWF and its derivation can be found in, \textit{e.g.}, Ref.~\cite{teufel2009bohmian}. We will then derive the non-linear Schr\"odinger equation (NSE) that dictates its evolution, a key ingredient in identifying the different contributions to the energy fluctuations.

\subsection{Definition}

In the non-relativistic Bohmian approach, a system is modeled by a collection of $ N $ point particles and a dynamical law for their motion is provided.
We consider Hamiltonians of the form $ \hat{H}(t) = \sum_{i=0}^{N-1} {\hat{P}^{(i)^2}}/(2m\ui) + V(\{\hat{Z}\ui \}, t) $, where $ \hat{P}\ui $ and $ \hat{Z}\ui $ are the three-dimensional momentum and position operators of particle $ i $, respectively, $ m\ui $ is the mass of particle $ i $ and $ V $ is a function of the operators $ \hat{Z}\ui $ and time only. In this case, the $ j $ Cartesian component of the velocity field of particle $ i $ is given by
\begin{equation}\label{eq:vi_cwf}
	v\ui_j(\vec{z}, t) = \frac{\hbar}{m^{(i)}}
	\frac
	{\proj{\acom{\hat{P}_j\ui}{\hat{\sigma}\up(t)}}{\vec{z}}}
	{\proj{\hat{\sigma}\up(t)}{\vec{z}}},
\end{equation}
where $ \hat{\sigma}\up(t) = \dyad{\Psi(t)} $ with $ \ket{\Psi(t)} $ the wave function of the combined system and environment, $ \{\cdot,\cdot \} $ is the anti-commutator,
$ \vec{z} := (\vec{z}^{(0)}, \dots, \vec{z}^{(N-1)}) $ is a point in the configuration space of the $ N $ particles with $ \vec{z}\ui $ the
three-dimensional position of particle $ i $, $ \hat{P}\ui_j $ is the momentum operator associated with the
$ j^\text{th} $ component of the momentum of particle $ i $ and $ \hat{\Pi}(\vec{z}) := \dyad*{\vec{z}} $,
where $ \ket*{\vec{z}} $ is the basis vector associated with the configuration point $ \vec{z} $.
Note that $ v\ui_j(\vec{z},t) $ is proportional to the real part of the weak value of the momentum operator $\hat{P}\ui_j$ with pre-selection
on $\ket{\Psi(t)}$ and post-selection on position~\cite{AAV88,Flack_2018,flack2014weak}. Integrating the above equation yields a formal expression for the trajectory of each particle as
\begin{equation}\label{eq:z_i}
	\vec{Z}\ui(t|\vec{z}_0) = \vec{Z}\ui(0|\vec{z}_0) + \int_0^t \vec{v}\ui \left(\vec{Z}(s|\vec{z}_0), s\right) \dd s,
\end{equation}
where $  Z\ui(0|\vec{z}_0) $ is the initial position of particle $ i $
and $ \vec{Z}(t|\vec{z}_0) $ is the configuration trajectory given the initial condition $ \vec{z}_0 $. Note that the velocity of each particle is implicitly dependent on the wave function
$ \ket*{\Psi(t)} $, as well as the position of all the other particles through Eq.~\eqref{eq:vi_cwf}.
Each run of an experiment corresponds to a different initial condition \( \vec{z}_0 \) sampled randomly from the initial distribution \( \mathbb{P}_0(\vec{z}_0) = \Tr*{\hat{\Pi}(\vec{z}_0) \hat\sigma\up(0)} \), which is just the Born rule.
Thus, each run corresponds to a different trajectory \( \vec{Z}(t|\vec{z}_0) \) and the velocity field Eq. \eqref{eq:vi_cwf} guarantees that the Born rule is obeyed at all subsequent times~\cite{holland95}.

Now let us take particle $ i = 0 $ as our system of interest. Hereafter, we refer to this particle as ``the system'' and all
other particles as ``the environment''.
The idea behind the CWF of the system is the following \textemdash\
given an initial configuration $ \vec{z}_0 $, the CWF generates exactly the same system trajectory that is generated by the combined system and environment wave function $ \ket{\Psi(t)} $ from Eqs.~\eqref{eq:vi_cwf} and \eqref{eq:z_i}.
If the same trajectories are generated, the same statistical predictions are obtained~\cite{teufel2009bohmian}.
The wave function $ \ket{\Psi(t)} $ lives on the Hilbert space associated with the $ N $ particles and evolves according to a linear Schr\"odinger equation (LSE).
The conditional wave function however lives on the reduced Hilbert space of the system only and evolves according to a nonlinear Schr\"odinger equation (NSE).
Of particular interest is that the NSE contains additional terms relative to the LSE which are directly related to entanglement.
This allows us to trace back the role of entanglement in the evolution of the conditional wave function and, ultimately, in the energy fluctuations of the system.

Let us denote the position of the system by $ \vec{x} := \vec{z}^{(0)} $ and the position of the environment by
$ \vec{y} := (\vec{z}^{(1)}, \dots, \vec{z}^{(N-1)}) $, such that we can write $ \vec{z} = (\vec{x}, \vec{y}) $. The
unnormalized CWF for the system is then defined as
\begin{align}\label{eq:cwf}
	\ket*{\ucwf(t|\vec{z}_0)} & = \int \dd\mathbf{x}\ \ucwf(\vec{x},t|\vec{z}_0) \ket{\vec{x}} \nonumber  \\
	                          & = \int \dd\mathbf{x}\ \Psi(\vec{x},\vec{Y}(t|\vec{z}_0),t) \ket{\vec{x}},
\end{align}
where $ \dd\mathbf{x} $ is an infinitesimal three-dimensional volume element,
$ \vec{Y}(t|\vec{z}_0) := (\vec{Z} ^{(1)}(t|\vec{z}_0), \dots, \vec{Z} ^{(N-1)}(t|\vec{z}_0)) $ is the trajectory of the environment configuration and the position representation of
$ \ket*{\Psi(t)} $ at the point $ \vec{z} = (\vec{x}, \vec{Y}(t|\vec{z}_0)) $ is
$ \Psi(\vec{x},\vec{Y}(t|\vec{z}_0),t) := \braket*{\vec{x}, \vec{Y}(t|\vec{z}_0)}{\Psi(t)} $.
The CWF is conditioned by the trajectory of the
environmental degrees of freedom and thus an explicit function of $\vec{x}$ and $t$ only. In this sense, it is a single-particle wave function.
Further, it admits an operational interpretation within standard quantum mechanics in terms of weak measurements.
The weak value of the projection operator $ \dyad{\vec{x}} $, when pre-selected on $ \ket{\Psi(t)} $ and post-selected on a system zero-momentum state and environment position $ \vec{y} = \vec{Y}(t|\vec{z}_0) $, is proportional to $ \Psi(\vec{x},\vec{Y}(t|\vec{z}_0),t) = \ucwf(\vec{x},t|\vec{z}_0) $. Thus, the CWF trajectories are experimentally accessible. For further details, see Ref. \cite{norsen14}.

\subsection{Dynamics}

To derive the NSE governing the CWF evolution we look at its explicit time derivative. Taking into account the explicit time dependence of $ \Psi(\vec{x},\vec{Y}(t|\vec{z}_0),t) $ through $ \vec{Y}_t := \vec{Y}(t|\vec{z}_0) $, we have,
\begin{align}\label{eq:dtketphi}
	 & \partial_t \ket*{\ucwf(t|\vec{z}_0)}
	= \int\dd\mathbf{x}\ \partial_t\Psi(\vec{x},\vec{Y}_t, t) \, \ket*{\vec{x}} \nonumber                                                                                                                    \\
	 & = -\frac{\imath}{\hbar}\int\dd\mathbf{x}\ \left[ \mel*{\vec{x}, \vec{Y}_t}{\hat{H}(t)}{\Psi(t)} - \vec{v}^{(y)} \cdot \mel*{\vec{x}, \vec{Y}_t}{\hat{\vec{P}}^{(y)}}{\Psi(t)} \right] \ket*{\vec{x}},
\end{align}
where we have
introduced the shorthand notation $\vec{v}^{(y)} \cdot \mel*{\vec{x}, \vec{Y}_t}{\hat{\vec{P}}^{(y)}}{\Psi(t)} := -\imath\hbar\sum_{i=1}^{N-1}\sum_{j=1}^{3} v\ui_j\left( \vec{Z}(t,\vec{z}_0), t \right)\ (\partial_{y\ui_j} \Psi)(\vec{x},\vec{Y}_t,t) $ with $ \partial_{y\ui_j} $ the partial derivative with respect to the $ j^\text{th} $ component of the
position of the environment particle $ i $. We have used the fact that, by definition, $ \dd_t Y\ui_j(t|\vec{z}_0) = v\ui_j \left(\vec{Z}(t|\vec{z}_0),t \right) $
and that $ (\partial_{y\ui_j} \Psi)(\vec{x},\vec{Y}_t,t) = (\imath/\hbar) \mel*{\vec{x}, \vec{Y}_t}{\hat{P}\ui_j}{\Psi(t)} $.

Let $ \hat{H}(t) = \HS(t) + \HE + \HI $, where $ \HS(t) $ is the system
Hamiltonian acting only on the system degrees of freedom, $ \HE $ is the environment
Hamiltonian acting only on the environment degrees of freedom and $ \HI $ is an interaction
Hamiltonian which acts on both system and environment degrees of freedom.
From the Schr\"odinger equation $ \imath\hbar\partial_t\ket{\Psi(t)} = \hat{H}(t)\ket{\Psi(t)} $, it then follows that the evolution of the
CWF in the position representation can be written in the {generalised Schr\"odinger} form
\begin{align}\label{eq:dtphi}
	 & \partial_t \ucwf(\vec{x},t|\vec{z}_0) = \braket*{\vec{x}, \vec{Y}_t}{\partial_t \Psi(t)} + \frac{\imath}{\hbar} \vec{v}^{(y)} \cdot \mel*{\vec{x}, \vec{Y}_t}{\hat{\vec{P}}^{(y)}}{\Psi(t)} \nonumber \\
	 & = -\frac{\imath}{\hbar} \left\{ \mel*{\vec{x}, \vec{Y}_t}{\hat{H}(t)}{\Psi(t)} -  \vec{v}^{(y)} \cdot \braket*{\vec{x}, \vec{Y}_t}{\hat{\vec{P}}^{(y)}\Psi(t)} \right\} \nonumber                     \\
	 & = -\frac{\imath}{\hbar} \left\{ \mel*{\vec{x}}{\HS(t)}{\tilde\phi(t|\vec{z}_0)} + \mel*{\vec{x}, \vec{Y}_t}{\HI}{\Psi(t)}   \right. \nonumber                                                         \\
	 & \qquad \quad \left. + \mel*{\vec{x}, \vec{Y}_t}{\HE}{\Psi(t)}  -  \vec{v}^{(y)} \cdot \mel*{\vec{x}, \vec{Y}_t}{\hat{\vec{P}}^{(y)}}{\Psi(t)} \right\}.
\end{align}
This form is particularly interesting because we can decompose the evolution of the CWF into three distinct terms.
The first term is the contribution from the system's own Hamiltonian.
The second term contains the effect of the interaction term.
The last two terms, perhaps the most interesting here, can be associated with entanglement between the system and environment (see Section \ref{sec:cond_energy} below).

Finally, for the purpose of calculations and presentation, it is more convenient to express
Eqs.~\eqref{eq:dtketphi} and~\eqref{eq:dtphi} in terms of the unnormalized density operator
$ \ucdm(t|\vec{z}_0) = \dyad*{\ucwf(t|\vec{z}_0)} $. The evolution of $ \ucdm $ is then given by
\begin{align}\label{eq:dtrho}
	\dd_t\ucdm(t|\vec{z}_0) = \dd_t{\hat{\tilde{\rho}}}\up_{\rm S}(t|\vec{z}_0)
	+ \dd_t{\hat{\tilde{\rho}}}\up_\text{int}(t|\vec{z}_0)
	+  \dd_t{\hat{\tilde{\rho}}}\up_\text{ent}(t|\vec{z}_0),
\end{align}
where
\begin{align}
	\dd_t{\hat{\tilde{\rho}}}\up_{\rm S} =      & -\frac{\imath}{\hbar} [\HS, \hat{\tilde\rho}\up] ,                                                    \\
	\dd_t{\hat{\tilde{\rho}}}\up_{\text{int}} = & -\frac{\imath}{\hbar}  \evy{[\HI, \hat{\sigma}\up(t)]} ,                                              \\
	\dd_t{\hat{\tilde{\rho}}}\up_\text{ent} =   & -\frac{\imath}{\hbar}
	\left[ \evy{[\HE, \hat{\sigma}\up(t)]} \right. \nonumber                                                                                            \\
	                                            & \left.- v^{(y)} \cdot \evy{[\hat{\vec{P}}^{(y)}, \hat{\sigma}\up(t)]} \right] . \label{eq:dtrhoterms}
\end{align}
We have here used the notation $ \evy{\hat{A}} := \projE{\hat{A}}{\vec{Y}_t} $
for any operator $ \hat{A} $ in the combined Hilbert space of system and environment where $ \hat{\Pi}^\text{E}(\vec{y}) := \dyad*{\vec{y}} $ is a
projection operator similar to $ \hat{\Pi}(\vec{z}) $ but acting only on the environment degrees of freedom.
We note that this equation is stochastic in nature despite having no explicit noise terms, because the initial conditions are
stochastic (at least for the position $\vec{z}_0$).
However, once $\vec{z}_0$ and $ \ket*{\Psi(0)} $ have been specified the evolution is completely deterministic.

\section{Conditional energy}\label{sec:cond_energy}

To study energy fluctuations we need first to provide a link between the CWF of the system and its energy. Different proposals can be found in the literature to study fluctuations in open quantum
systems~\cite{Sampaio18,Elouard17,Miller2016,Hekking:2013aa,Horowitz2012,Solinas2013,Talkner2016,Campisi2009}.
For descriptions based on stochastic states (pure or mixed), a typical approach to estimate the energy of the system is to take the
expectation value of the system Hamiltonian with respect to the stochastic state (see, \textit{e.g.}, Refs.~\cite{Elouard17,Alonso2016}). We adopt this approach here. To this end, we introduce the concept of {\it conditional energy} as the expectation value of $ \HS(t) $ with respect to the CWF. Explicitly, we define the quantity
\begin{align}\label{eq:condenergy}
	u\up(t|\vec{z}_0) & := \frac{\ev*{\HS(t)}{\ucwf(t|\vec{z}_0)}}{\braket{\ucwf(t|\vec{z}_0)}} \nonumber                \\
	                  & = \frac{\Tr{\hat{\tilde\rho}\up(t|\vec{z}_0) \, \HS(t)}}{\Tr{\hat{\tilde\rho}\up(t|\vec{z}_0)}},
\end{align}
as the \textit{conditional energy} of the system.
In analogy to the CWF, the conditional energy is conditioned by a trajectory of the environment, or equivalently, by the initial condition $ \vec{z}_0 $.
We will demonstrate below that this conditional energy provides a meaningful way to estimate the energy of the
system for individual trajectories of the environment. To identify the different terms contributing to the fluctuations of the
conditional energy we look at its total time derivative, namely
\begin{align} \label{eq:energyflow}
	\dd_t u\up(t|\vec{z}_0)
	= & \frac{\Tr{ \left( \HS(t) - u\up(t|\vec{z}_0) \right) \, \dd_t\ucdm(t|\vec{z}_0)}}
	{\Tr{\hat{\tilde\rho}\up(t|\vec{z}_0)}} \nonumber                                                          \\
	  & + \frac{\Tr{\hat{\tilde\rho}\up \, \dd_t{\hat{H}}_{\rm S}(t)}}{\Tr{\hat{\tilde\rho}\up(t|\vec{z}_0)}}.
\end{align}
Using Eq.~\eqref{eq:dtrho} it is straightforward to show that the conditional energy flow splits into three terms as
\begin{align}\label{eq:du}
	\dd_t u\up(t|\vec{z}_0) = \dd_t{u}\up_{\text{ext}}(t|\vec{z}_0) + \dd_t{u}\up_{\text{int}}(t|\vec{z}_0)
	+ \dd_t{u}\up_{\text{ent}}(t|\vec{z}_0),
\end{align}
where
\begin{align}
	\dd_t{u}\up_\text{ext}(t|\vec{z}_0)
	= & \frac{\Tr{\hat{\tilde\rho}\up(t|\vec{z}_0) \, \dd_t{\hat{H}}_{\rm S}(t)}}{\Tr{\hat{\tilde\rho}\up(t|\vec{z}_0)}} ,   \label{eq:du_exp}                                        \\
	\dd_t{u}\up_{\text{int}}(t|\vec{z}_0)
	= & \frac{\Tr{ \left(\HS(t) - u\up(t|\vec{z}_0) \right) \, \dd_t{\hat{\tilde\rho}}\up_{\text{int}}(t|\vec{z}_0)}}{\Tr{\hat{\tilde\rho}\up(t|\vec{z}_0)}} , \label{eq:du_int} \\
	\dd_t{u}\up_{\text{ent}}(t|\vec{z}_0)
	= & \frac{\Tr{\left(\HS(t) - u\up(t|\vec{z}_0) \right) \, \dd_t{\hat{\tilde\rho}}\up_{\text{ent}}(t|\vec{z}_0)}}{\Tr{\hat{\tilde\rho}\up(t|z_0)}} . \label{eq:du_ent}
\end{align}
This is our main result and allows a straightforward physical interpretation according to which the three terms can be called the {\it external, interaction and entanglement} contributions to the rate of energy exchange, respectively. 
This is based on the following observations:
(i) If there is no explicit time dependence in the system Hamiltonian, i.e. $\dd_t{\hat{H}}_{\rm S} = 0$,  the first term vanishes, i.e.  $\dd_t{u}_\text{ext} = 0$, while
(ii) in the absence of interaction between the system and the environment, i.e. $\HI = 0$, the second term vanishes, i.e. $\dd_t u\up_{\text{int}} = 0$.
Finally,
(iii) if the system and environment are not entangled at time $t$, the entanglement contribution to the conditional energy flow vanishes, i.e. $\dd_t u\up_{\text{ent}} = 0$.
To see this, consider a factorized state $\ket{\Psi(t)} = \ket{\phi(t)}\otimes \ket{\chi(t)}$, with $\ket{\phi(t)}$  the system wavefunction and $\ket{\chi(t)}$ the environmental wavefunction. In Eq.~\eqref{eq:dtrhoterms} the operators $\HE$ and $\hat{\vec{P}}^{(y)}$ do not act on the system, and hence the trace over the environment leaves just a term proportional to the system state, i.e. $\dd_t{\hat{\tilde\rho}}^{(\phi\otimes\chi)}_{\text{ent}}(t|\vec{z}_0) = f(t|\vec{z}_0) \,  \dyad{\phi(t)}$ with $ f(t|\vec{z}_0) $ a complex-valued function that depends only on time and the initial conditions $ \vec{z}_0 $.
Noting that the conditional wave function reduces to \( \ket*{\tilde\phi^{(\phi\otimes\chi)}(t|\vec{z}_0)} = \braket*{\vec{Y}_t}{\chi(t)}\ket{\phi(t)} \), we see that \( \Tr\{\hat{\tilde\rho}^{(\phi\otimes\chi)}(t|\vec{z}_0)\} = |\braket*{\vec{Y}_t}{\chi(t)}|^2 \) and the conditional energy reduces to \( u^{(\phi\otimes\chi)}(t|\vec{z}_0) = |\braket*{\vec{Y}_t}{\chi(t)}|^2\ev*{\HS(t)}{\phi(t)}/|\braket*{\vec{Y}_t}{\chi(t)}|^2 = \ev*{\HS(t)}{\phi(t)} \). Thus, if follows from Eq.~\eqref{eq:du_ent} that the entanglement contribution vanishes as,
	\begin{align}
		 & \dd_t{u}^{(\phi\otimes\chi)}_{\text{ent}}(t|\vec{z}_0) \nonumber                                                                      \\
		 & = \frac{f(t|\vec{z}_0)}{|\braket*{\vec{Y}_t}{\chi(t)}|^2}\left(\ev*{\HS(t)}{\phi(t)} - u^{(\phi\otimes\chi)}(t|\vec{z}_0)\right) = 0.
	\end{align}

To make the connection to the canonical average system energy \( \Tr{\HS(t) \, \hat\sigma\up(t)} \), we need to consider the statistical average of the conditional energy, $ \evv*{u\up(t)} $, over the initial configuration space points $ \vec{z}_0 $, namely,
\begin{align}\label{eq:ev-evv}
	 & \evv*{u\up(t)} = \int\dd \mathbf{z}_0 \abs*{\Psi(\vec{z}_0,0)}^2 u\up(t|\vec{z}_0) \nonumber                                                                                                                                                                  \\
	 & = \int\dd \mathbf{z}_0 \abs*{\Psi(\vec{z}_0,0)}^2 \ev*{\HS(t)}{\cwf(t|\vec{z}_0)} \nonumber                                                                                                                                                                   \\
	 & = \int\dd \mathbf{z}_0 \abs*{\Psi(\vec{z}_0,0)}^2 \frac{\ev*{\HS(t)}{\ucwf(t|\vec{z}_0)}}{\braket*{\ucwf(t|\vec{z}_0)}} \nonumber                                                                                                                             \\
	 & = \int\dd \mathbf{z}_0 \abs*{\Psi(\vec{z}_0,0)}^2 \nonumber                                                                                                                                                                                                   \\
	 & \quad \times \frac{\int\dd \mathbf{x}\braket*{\Psi(t)}{\vec{x},\vec{Y}_t}\mel*{\vec{x}, \vec{Y}_t}{\HS(t)}{\Psi(t)}}{\int\dd \mathbf{x}'\braket*{\Psi(t)}{\vec{x}',\vec{Y}_t}\braket*{\vec{x}', \vec{Y}_t}{\Psi(t)}} \nonumber                                \\
	 & = \int\dd \mathbf{y} \left[\int \dd \mathbf{x}' \abs*{\Psi(\vec{x}',\vec{y},t)}^2 \right] \frac{\int\dd \mathbf{x}\braket*{\Psi(t)}{\vec{x},\vec{y}}\mel*{\vec{x},\vec{y}}{\HS(t)}{\Psi(t)}}{\int\dd \mathbf{x}' \abs*{\Psi(\vec{x}',\vec{y},t)}^2} \nonumber \\
	 & = \int\dd \mathbf{x}\dd \mathbf{y} \braket{\Psi(t)}{\vec{x},\vec{y}}\mel{\vec{x},\vec{y}}{\HS(t)}{\Psi(t)} \nonumber                                                                                                                                          \\
	 & = \ev*{\HS(t)}{\Psi(t)} = \Tr{\HS(t) \, \hat{\sigma}\up(t)},
\end{align}
where we have used the equivariance property
$ \int \dd \mathbf{z}_0\abs*{\Psi(\vec{z}_0,0)}^2 f(Z\up(t,\vec{z}_0),t) = \int \dd \mathbf{z}\abs*{\Psi(\vec{z},t)}^2 f(\vec{z},t) $,
valid for any $L^2$ function $ f $, and $ \dd\mathbf{z} = \dd\mathbf{x}\dd\mathbf{y} $ where $\dd\mathbf{y}$ is an infinitesimal volume in the
configuration space of the environment. This shows that the statistical average of the conditional energy over the initial configurations indeed gives the physically meaningful expectation value of the system Hamiltonian $\HS(t)$.

Furthermore, we can relate the statistical average of the conditional energy flow (Eqs.~\eqref{eq:du} to \eqref{eq:du_ent})
to the average energy flow $ \dd_t \ev{\HS(t)}{\Psi(t)} =  \Tr \{\hat{\sigma}\up(t) \dd_t\HS(t)\} + \Tr \{\HS(t) \dd_t\hat{\sigma}\up(t)\} $. For the first term, we find
\begin{equation}\label{eq:exener}
	\Tr{\hat\sigma\up(t) \, \dd_t\HS(t)} = \evv*{\dd_t u\up_\text{ext}(t)},
\end{equation}
which nicely confirms that energy exchange of the system that results from
time dependence of the Hamiltonian is given by our external contribution $\evv*{\dd_t u\up_\text{ext}(t)}$
to the total conditional energy. The second term, typically associated with entropy or heat exchange \cite{Vinjanampathy16}, is more
involved and is given by
\begin{align} \label{eq:entener}
	\Tr{\HS(t)\dd_t{\hat{\sigma}\up}(t)} = \evv*{\dd_t u\up_\text{int}(t)} + \evv*{\dd_t u\up_\text{ent}(t)}
\end{align}
where $ \evv*{\dd_t u\up_\text{ent}}(t) = \int\dd \mathbf{z}\ u\up(t,\vec{y})\ \ev*{\dd_t\hat{\tilde\rho}\up_\text{int}}{\vec{x}} $ and we have defined $ u\up(t,\vec{y}) $ to be the conditional energy at time $t$ given a configuration $\vec{y}$ of the environment at the same time.
Defining the state-dependent operator $ \hat{u}\up(t) $ acting only on the environment degrees of freedom with matrix elements
$\mel{i}{\hat{u}\up(t)}{j} = \int\dd \mathbf{y} \braket{i}{\vec{y}}\braket{\vec{y}}{j} u\up(t,\vec{y})$, the entanglement contribution to the
expectation value of the energy can be written in the compact form
\begin{align}\label{eq:evvdotent}
	\evv*{\dd_t{u}\up_\text{ent}(t)} = -\frac{\imath}{\hbar}\Tr{\hat{\sigma}\up(t) [\hat{u}\up(t), \HI]}.
\end{align}
This form shows that if no interaction is present or the interaction Hamiltonian commutes with $\hat{u}\up(t)$,
entanglement makes no direct contribution to the average energy change. This is in accordance with the expectation that on average no energy transfer can come about from entanglement alone, {\it i.e}, an interaction with the environment is required.
We give an example below using a momentum-momentum interaction where this term is relevant.

We conclude this section with two remarks. First, the energy exchange contributions identified here are valid for any environment and interaction Hamiltonians, including time-dependent interactions. In particular, no assumption is made about the structure of the environment and the form or strength of the interaction. We note that we have analysed energy fluctuations associated with the bare system Hamiltonian $\HS(t)$, i.e., the fluctuations of the term $ u\up(t|\vec{z}_0) = \ev*{\HS(t)}{\cwf(t|\vec{z}_0)}$. When the system-environment interaction is strong, alternative system Hamiltonians have been investigated, especially in the thermodynamic setting~\cite{philbin16, seifert16, jarzynski17,strasberg17,Miller17,aurell18,miller18}, which attribute part of the interaction energy to the system energy. Additional energetic fluctuations could arise in this effective picture due to the difference between the bare and effective Hamiltonians.

Second, it is important to note that in deriving the results in this section we have assumed that the state $ \ket{\Psi(t)} $ is known. In many experimental scenarios, the system is prepared in some well-known state before each run of a given drive protocol and if this state is known, quantities conditioned on the individual pure states are empirically accessible.
However, if one \textit{only} has access to the averages described through the relevant density operator $\hat{\sigma}$ representing a mixed state,
all measurable quantities have to be expressed at this level and they cannot depend on an individual decomposition of $\hat{\sigma}$.
If they did, one could distinguish between different decomposition of $\hat{\sigma}$ without \textit{a priori} knowledge of its preparation, in contradiction with the predictions of quantum mechanics.
We show in the Appendix how Eqs.~\eqref{eq:du} to \eqref{eq:evvdotent} can be coarse-grained to describe general mixed states based on Refs.~\cite{norsen14,Durr05}. The main result remains unaltered, with the only difference that the entanglement contribution should now be considered as a correlations contribution,since it will be non-zero for both quantum and classical correlations (see Ref.~\cite{Alipour16} for a similar structure using reduced density operators).

\section{Examples}\label{sec:examples}

In this section we explicitly solve a series of simple examples to show how CWFs can be used to characterize energy flow in
open quantum systems beyond the usual expectation values of the Hamiltonian.

\subsection{Driven {system} without interaction or entanglement}

The simplest example is that of no interaction $\HI = 0$ and no entanglement between the system and the environment.
For this case the only contribution to energy change should come from the external drive term $ \dd_t{u}\up_{\text{ext}} $.
We can then write the total system state as a product state between the system $\hat{\rho}$ and the environment $\hat{\chi}$ as $\hat{\sigma}\up(t) = \hat{\rho}(t) \otimes \hat{\chi}(t)$.
Noting that $ \dd_t{\hat{\tilde\rho}}\up_\text{ent} \propto \hat{\rho}(t) $, we see that $ \dd_t{u}\up_{\text{ent}} \propto \Tr{\hat{\rho} \HS} - u\up = 0 $. Thus, as expected, the total rate of energy change is given by $ \dd_t u\up = \dd_t{u}\up_\text{ext} $.

\subsection{Driven {system} with interaction but no entanglement}

Next, we consider the case of a driven system which interacts with the environment but is not entangled with it at any time, \textit{i.e.}, $ \ket{\Psi(t)} = \ket{\phi(t)}\otimes\ket{\chi(t)} $.
As an example, consider the case where the interaction has negligible effect on the state of the environment.
Let the total system be the product state ${\hat\sigma}\up(t) = \hat{\rho}(t) \otimes \hat{\chi}(t)$,
where $ \rho(t) = \dyad{\phi(t)} $ and $ \hat{\chi}(t) = \dyad{\chi(t)} $, and an interaction
Hamiltonian of the form $\HI(t) = \sum_i \hat{A}_i(t) \otimes \hat{B}_i(t)$.
Then
\begin{align}
	\dd_t{\hat{\tilde\rho}}\up_\text{int}(t|\vec{z}_0)
	 & = -\frac{\imath}{\hbar}\evy{[\HI,\hat{\sigma}\up(t)]} \nonumber                                \\
	 & = -\frac{\imath}{\hbar} \sum_i [\hat{A}_i \hat{\rho}(t)\evy{\hat{B}_i \hat{\chi}(t)} \nonumber \\
	 & \quad - \hat{\rho}(t)\hat{A}_i \evy{ \hat{\chi}(t) \hat{B}_i}]
\end{align}
If $  \evy{\hat{B}_i \hat{\chi}} \approx \evy{ \hat{\chi} \hat{B}_i} := b_i(t|\vec{z}_0) \ \forall i,Y_t $, then
$ \dd_t{\hat{\tilde\rho}}_\text{int} \approx - (\imath / \hbar) \sum_i b_i(t|\vec{z}_0) [\hat{A}_i(t), \hat{\rho}(t)] $.
Furthermore, the state remains factorized and it is given by
\begin{equation}
	\ket{\phi(t+\delta t)} = \ket{\phi(t)} - \frac{\imath \delta t}{\hbar}\left[\HS(t) + \sum_i b_i(t|\vec{z}_0) \hat{A}_i \right] \ket{\phi(t)},
\end{equation}
to first order in $\delta t$. Thus, the system is effectively driven by an additional term $\hat{H}_\text{D}(t|\vec{z}_0) \equiv  \sum_i b_i(t|\vec{z}_0) {\hat A}_i(t)$.

Similarly to the previous subsection the entanglement contribution vanishes since $ \dd_t{\hat{\tilde\rho}}\up_\text{ent} \propto \hat{\rho}(t) $.
For the interaction term we have
\begin{align}
	\dd_t{u}\up_\text{int} = & -\frac{\imath}{\hbar} \Tr{\sum_i b_i(t|\vec{z}_0) \, \left[\hat{A}_i(t), \, \hat{\rho}(t)\right] \, \left(\HS(t) - u\up(t|Y_t)\right)} \nonumber \\
	=                        & -\frac{\imath}{\hbar}\Tr{\hat{\rho}(t) \left[\HS(t), \, \sum_i b_i(t|\vec{z}_0) \hat{A}_i(t) \right]} \nonumber                                  \\
	=                        & -\frac{\imath}{\hbar}\Tr{\hat{\rho}(t) \left[\HS(t), \, \hat{H}_\text{D}(t|\vec{z}_0) \right]}. \label{eq:u_int_laser}
\end{align}

A concrete example here is that of a qubit driven by a laser.
This is described by an interaction with a harmonic oscillator in a coherent state $ \ket{\alpha} $ ($ \abs{\alpha}^2 \gg 1 $) with an interaction
Hamiltonian (neglecting fast oscillating terms) $ \HI = \lambda_{\rm c} \exp(i\omega t) \hat{a}^\dagger \hat{\sigma}^-  + \lambda_{\rm c}^*\exp(-i\omega t)
	\hat{a} \hat{\sigma}^+ $, where $ \omega $ is the natural frequency of the qubit, $ \lambda_{\rm c} $ is a coupling constant, $ \hat{a} $ ($ \hat{a}^\dagger $) the
annihilation (creation) operator, and $ \hat{\sigma}^- $ ($ \hat{\sigma}^+ $) the lowering (raising) operator. Let $ \hat\sigma\up(t) = \hat{\alpha}\otimes\hat{\rho} $ be the initial state where $ \hat{\alpha} $ is the density operator of a coherent state $ \ket{\alpha} $.
Making the approximation that $ \hat{a}^\dagger \ket{\alpha} \approx \alpha^*\ket{\alpha} $ leads to
$ \hat{H}_{\rm D} = (c \hat{\sigma}^- + c^* \hat{\sigma}^+ ) $  in Eq. \eqref{eq:u_int_laser}, where $ c = \lambda_{\rm c} \alpha \exp(i\omega t) $. Thus,
\begin{equation}
	\dd_t{u}_\text{int}^\text{laser} = -\frac{\imath}{\hbar} \Tr{\hat{\rho}(t) \, \left[\HS(t), \, c \hat{\sigma}^- + c^* \hat{\sigma}^+ \right]},
\end{equation}
as expected. Note that if we were to include the laser interaction as part of the system energy, this contribution would appear as an explicit time dependent term.

\subsection{Driven {system} with interaction and entanglement}\label{sec:ex3}

\begin{figure}
	\includegraphics[width=1\linewidth]{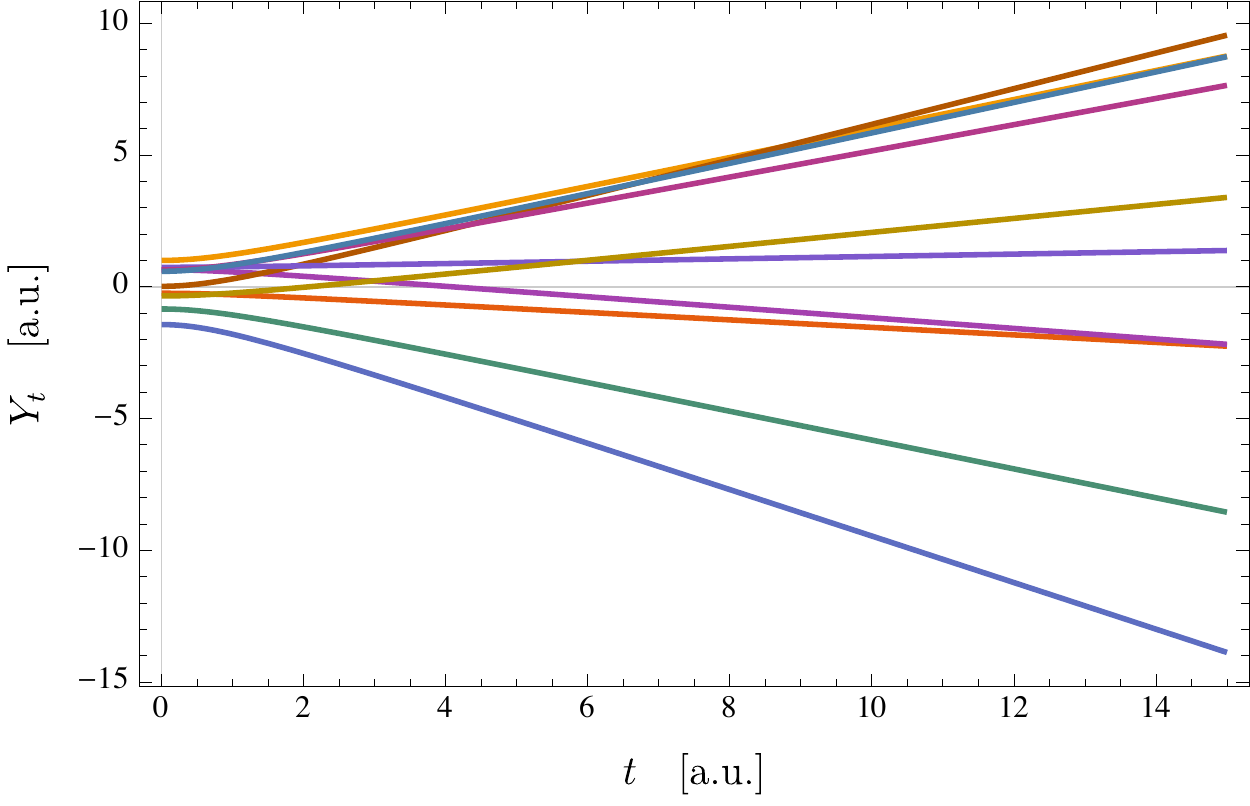}
	\caption{Typical trajectories $Y_t$ for the case of two particles interacting through a quadratic potential in Sec. \ref{sec:ex3}
		with initial conditions randomly sampled from the probability distribution $ \abs{\Psi(x,y,0)}^2 $. The colors indicate different
		initial conditions.}
	\label{fig:boson_traj_y}
\end{figure}

Here we consider two different cases where both interaction and entanglement contributions are present.

\subsubsection{$\hat{X}\otimes\hat{Y}$ interaction}\label{sec:ex3.1}

For the sake of simplicity we consider here the case of two particles interacting through a quadratic potential.
The Hamiltonian in units of $ m = \hbar = 1 $ is given by
\begin{equation}\label{eq:H_int_ent}
	\hat{H} = \frac{1}{2} ( \hat{P}_X^2 + \hat{P}_Y^2 ) + \frac14(\hat{X} - \hat{Y})^2,
\end{equation}
and we start from an initial factorized state $ \Psi(x,y,0) = \pi^{-1/2}\exp[-(x^2+y^2)/2] $.
Solving the Schr\"odinger equation we explicitly find that the full wave function evolves as
\begin{equation}
	\Psi(x,y,t) = \frac{1}{\sqrt{\pi(1+\imath t)}}e^{-\frac14\left[(x+y)^2 + (x-y)^2/(1+\imath t) + 2\imath t \right]},
\end{equation}
and for the trajectories we get
\begin{align}
	Y_t & := Y(t|x_0,y_0) = b(t) x_0 + a(t) y_0,
\end{align}
with $ a(t) = (\sqrt{1+t^2} + 1)/2 $ and $ b(t) = (\sqrt{1+t^2} - 1)/2 $. Since there are only two particles in one dimension we will drop the indices
$ i $ and $ j $ and use the notation $ x_0 $ and $ y_0 $ to denote the initial conditions of the $ X $ and $ Y $ particles.
Figure \ref{fig:boson_traj_y} shows some of the environment trajectories with initial
conditions sampled randomly from the probability distribution $ \abs{\Psi(x,y,0)}^2 $.
The unnormalized CWF is explicitly given by
\begin{align}\label{eq:psi_int_ent}
	 & \tilde\phi(x,t|x_0,y_0) \nonumber                                                                                                           \\
	 & =  \frac{\exp{-2\imath t}}{\sqrt{\pi(1+\imath t)}}e^{-\frac{1}{4} \{[x+b(t)x_0 + a(t)y_0]^2 + [x-b(t)x_0 - a(t)y_0]^2/(1+\imath t)\}}.
\end{align}
We take the system Hamiltonian to be $ \HS = \hat{P}_X^2/2 + \hat{X}^2/4 $ and the interaction Hamiltonian $ \HI = -\hat{X} \hat{Y}/2$.
The conditional energy and its time derivative can be readily obtained as
\begin{equation}
	u\up(t|x_0,y_0) = \frac{3}{8} + \frac{t^2 Y_t^2}{4 t^2+8},
\end{equation}
and
\begin{equation}
	\dd_t{u}\up(t|x_0,y_0) = \frac{t Y_t \left(t \left(t^2+2\right) \dd_t Y_t +2 Y_t\right)}{2 \left(t^2+2\right)^2}.
\end{equation}

\begin{figure}
\includegraphics[width=1\linewidth]{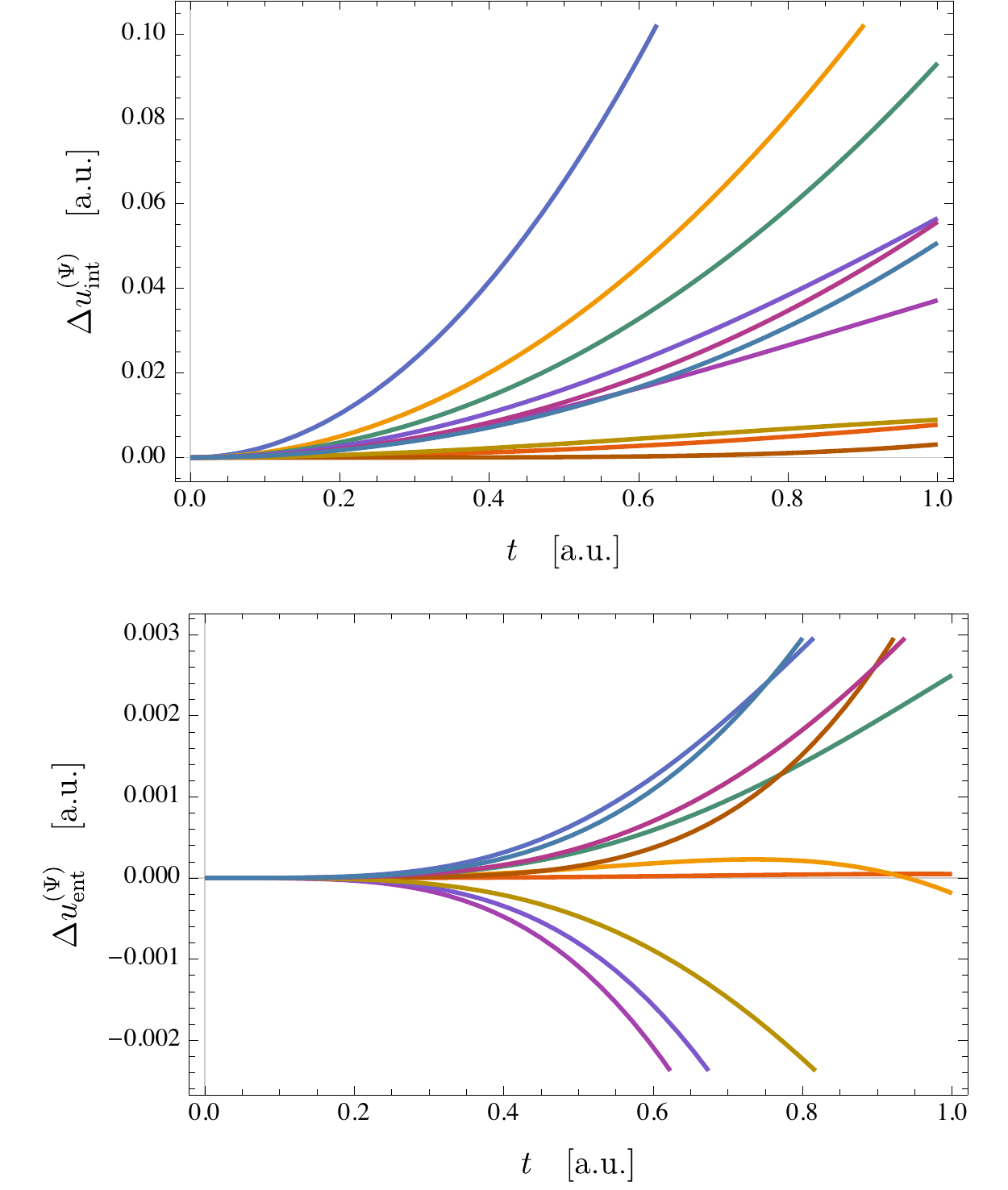}
\caption{Time evolution of the energy components (a) $ \Delta{u}_\text{int} $ and (b) $ \Delta{u}_\text{ent} $ corresponding to the trajectories
	$Y_t$ in Fig. \ref{fig:boson_traj_y}. The colors indicate different initial conditions as in Fig. \ref{fig:boson_traj_y}.}
\label{fig:boson_u_int_ent}
\end{figure}

Similarly, the interaction component of $ \dd_t{u} $ is given by inserting Eq. \eqref{eq:psi_int_ent} in  Eq. \eqref{eq:du_int} which yields
\begin{equation}
	\dd_t{u}\up_\text{int}(t|x_0,y_0) = \frac{t Y_t^2}{2(2 + t^2)},
\end{equation}
and the entanglement contribution is then determined by $ \dd_t{u}_\text{ent} = \dd_t{u} - \dd_t{u}_\text{int} $, which yields
	\begin{align}
		\dd_t{u}\up_\text{ent}(t|x_0,y_0) = \frac{t^2 Y_t \left(\left(t^2+2\right) \dd_t Y_t - t Y_t\right)}{2 \left(t^2+2\right)^2}.
	\end{align}
The total contributions from the interaction and entanglement can be written as
\begin{align}
	 & \Delta u\up_\text{int}(t|x_0,y_0) = \int_0^t \dd_t{u}\up_\text{int}(t|\vec{z}_0) \dd t = \nonumber \\
	 & = \frac{1}{16} \left[ 4 \left(x_0^2-y_0^2\right) \arctan{\sqrt{t^2+1}}  \right. \nonumber     \\
	 & \left. -(x_0+y_0) [c(t) x_0- d(t) y_0] + 4 x_0 y_0 \log \frac{2}{t^2+2}\right];                    \\
	 & \Delta u\up_\text{ent}(t|x_0,y_0) = \Delta u\up(t|x_0,y_0) - \Delta u\up_\text{int}(t|x_0,y_0),
\end{align}
respectively, where $c(t) = -t^2+4 \sqrt{t^2+1}+\pi -4 $, $ d(t)=t^2+4 \sqrt{t^2+1}+\pi -4 $ and $ \Delta u\up(t|x_0,y_0) = u\up(t|x_0,y_0) - u\up(0|x_0,y_0) $.
Figure \ref{fig:boson_u_int_ent} shows the evolution of $ \Delta u_\text{int} $ and $ \Delta u_\text{ent} $ for the trajectories $Y_t$ in
Fig. \ref{fig:boson_traj_y}.

As advertised in Sec. \ref{sec:cond_energy}, the average energy flow coming from the entanglement vanishes, i.e.
$ \evv*{\Delta u\up_\text{ent}} = 0 $, and $ \evv*{\Delta u\up} = \evv*{\Delta u\up_\text{int}} = t^2/16 $.
However, higher moments of the entanglement contribution do not vanish as evidenced by Fig.~\ref{fig:boson_u_int_ent} (b).
To give an estimate of how much the entanglement fluctuations contribute to the conditional energy fluctuations,
we can decompose the variance of the conditional energy into its entanglement and interaction contributions, namely,
\begin{align}
	\text{Var}\{\Delta u\up\} = & \text{Var}\{\Delta u\up_\text{int}\} + \text{Var}\{\Delta u\up_\text{ent}\} \nonumber \\
	& + 2 \text{Cov}\{\Delta u\up_\text{int},\Delta u\up_\text{ent}\},
\end{align}
where $ \text{Cov}\{z,w\} = \evv{zw} - \evv{z}\evv{w} $ is the covariance of $ z $ and $ w $ and $ \text{Var}\{z\} = \text{Cov}\{ z , z \} $ is the variance of $ z $.
Figure \ref{fig:boson_var} shows these quantities as a function of time. As already seen from Fig. \ref{fig:boson_u_int_ent}, the interaction
contribution quickly dominates over the entanglement one.

\begin{figure}
	\includegraphics[width=1\linewidth]{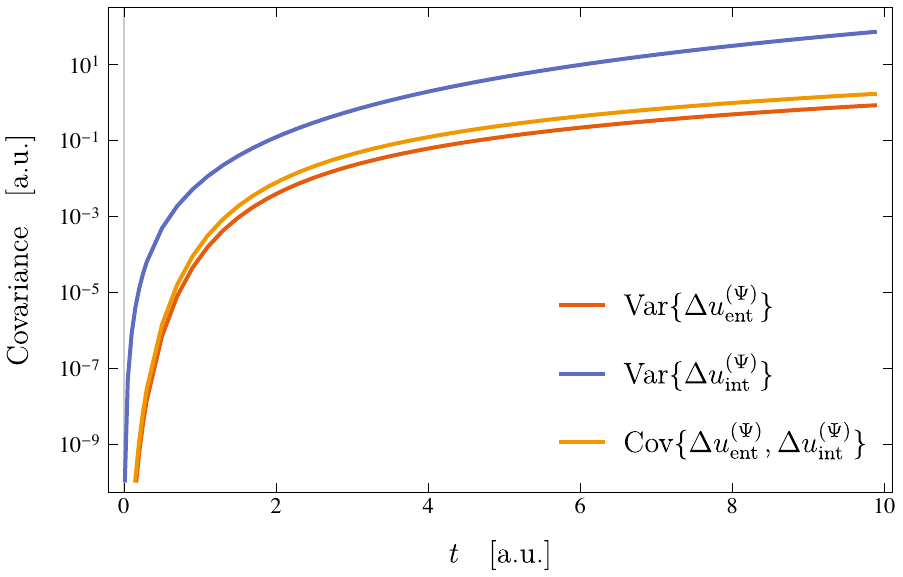}
	\caption{Variance of $ \Delta u $ decomposed into its entanglement and interaction contributions on a semilogarithmic scale (the first point is taken at $ t = 0.01 $). See text for details.}
	\label{fig:boson_var}
\end{figure}

\subsubsection{$\hat{P}_X\otimes\hat{P}_Y$ interaction}\label{sec:pxpy}

We consider now the case
\begin{equation}
	\hat{H}_\text{int} = - \lambda \hat{P}_X\otimes\hat{P}_Y,
\end{equation}
for some real-valued coupling constant $\lambda$.
This case is very similar to the previous one but now, as we will demonstrate, there is a nonzero entanglement contribution to the average energy flow.
To further simplify the analysis we assume $ \lambda \gg 1 $ such that the interaction term dominates over the system and environment
Hamiltonians and the evolution is dictated by $ \HI $. Again we assume an initial factorized state as in the previous case.
The solution of the Schr\"{o}dinger equation is then given by
\begin{equation}
	\Psi(x,y,t) = \frac{1}{\sqrt{\pi f(t)}} e^{-(x^2 + y^2 - 2\imath\lambda x y t)/2f(t)},
\end{equation}
with $f(t) = 1 + \lambda^2 t^2$. Rather than solving for the trajectories we will focus on the expectation value of the system Hamiltonian,
which in this case we take it to be just the kinetic energy, $\HS = \hat{P}_X^2/2$. A quick calculation shows that $\ev*{\HS}{\Psi(t)} = \frac{1}{4}$.
As expected, the average energy is constant since $ \HS $ commutes with $ \HI $. To evaluate the entanglement contribution from
Eq.~\eqref{eq:evvdotent} we need the conditional energy at time $ t $ given a configuration $ y $ of the environment, $ u\up(t, y) $, and the term $\ev{\dd_t{\hat{\tilde{\rho}}}_\text{int}}{x} $. The former is easily evaluated as before and gives
\begin{equation}\label{eq:u(y,t)_pxpy}
	u\up(t, y) =\frac{\lambda^2t^2 \left(2 y^2+1\right)+1}{4 f^2(t)}.
\end{equation}
The latter can be written in the form
\begin{align}\label{eq:rhointdot_pxpy}
	\ev{\dd_t{\hat{\tilde{\rho}}}_\text{int}}{x} & = -\imath \mel*{y,x}{[\hat{H}_\text{int},\hat{\sigma}\up]}{x,y} \nonumber \\
	& = \imath ( \Psi^*\partial_x\partial_y \Psi -  \Psi \partial_x\partial_y \Psi^* ) \nonumber \\
	& = -2 \Im ( \Psi^*\partial_x\partial_y \Psi ) \nonumber \\
	& =  \frac{2 \lambda  t}{\pi  f^3(t)} e^{-(x^2+y^2)/f(t)} [x^2+y^2-f(t)].
\end{align}
Finally, plugging Eqs.~\eqref{eq:u(y,t)_pxpy} and~\eqref{eq:rhointdot_pxpy} into Eq.~\eqref{eq:evvdotent} and performing the integration we find
\begin{equation}
	\evv*{\dd_t{u}\up_\text{ent}(t)} = \frac{\lambda ^3 t^3}{2 f^2(t)}.
\end{equation}
Since $\evv*{u\up(t)}$ is constant and there are no explicit time dependent terms in the Hamiltonian, it follows that
$\evv*{\dd_t{u}\up_\text{int}(t)} = -\evv*{\dd_t{u}\up_\text{ent}(t)}$. The entanglement and interaction flow contributions to the
expectation value of $\HS$ are plotted in Fig. \ref{fig:pxpy}.

\begin{figure}
	\centering
	\includegraphics[width=1\linewidth]{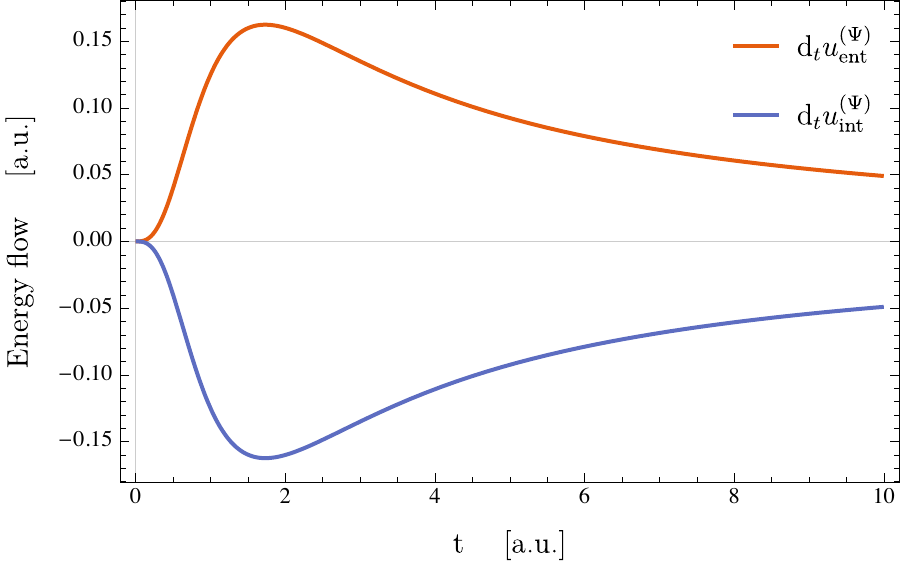}
	\caption{Entanglement and interaction flow contributions to the expectation value of $\HS$ for the example in section \ref{sec:pxpy}.}
	\label{fig:pxpy}
\end{figure}

\subsection{Driven {system} with entanglement but no interaction}\label{sec:ex4}

For the case of an entangled system with no interaction term, we consider two particles with spin $1/2$ initially in an entangled state
\begin{align}
	\Psi(x,y,0) = \frac{1}{\sqrt2}
	\begin{pmatrix}
		g(y) e^{\imath k x} f(x) \\
		g(y) f(x)                \\
		0                        \\
		0
	\end{pmatrix},
\end{align}
where $ f(x)=(2\pi\sigma_X^2)^{-1/4}\exp(-x^2/4\sigma_X^2) $ and $ g(y)=(2\pi\sigma_Y^2)^{-1/4}\exp(-y^2/4\sigma_Y^2) $
with $\sigma_X, \sigma_Y > 0$, and the spin components are represented in the basis
$\{ \ket{\uparrow_Y}\ket{\uparrow_X},   \ket{\downarrow_Y}\ket{\downarrow_X},  \ket{\uparrow_Y}\ket{\downarrow_X},  \ket{\downarrow_Y}\ket{\uparrow_X}\}$.
A unitary rotation $ \hat{U}_Y(t) = \exp(-\imath v t \hat{P}_Y \dyad{\uparrow_Y}) \otimes \mathds{1}_{X} $
acting {\it only} on particle $ Y $ is then applied,  where $ v $ is a real-valued constant, $t$ is the duration of the unitary and
$ \hat{P}_Y $ is the momentum operator
of particle $ Y $. We assume that $ t $ is short enough that the free evolution contribution from the individual particles can be ignored.
The resulting state is given by
\begin{align}\label{eq:rd_psi}
	\Psi(x, y, t) = \frac{1}{\sqrt2}\begin{pmatrix}
		g(y - vt) e^{ikx}f(x) \\
		g(y) f(x)             \\
		0                     \\
		0
	\end{pmatrix}.
\end{align}

\begin{figure}[t]
	\includegraphics[width=1\linewidth]{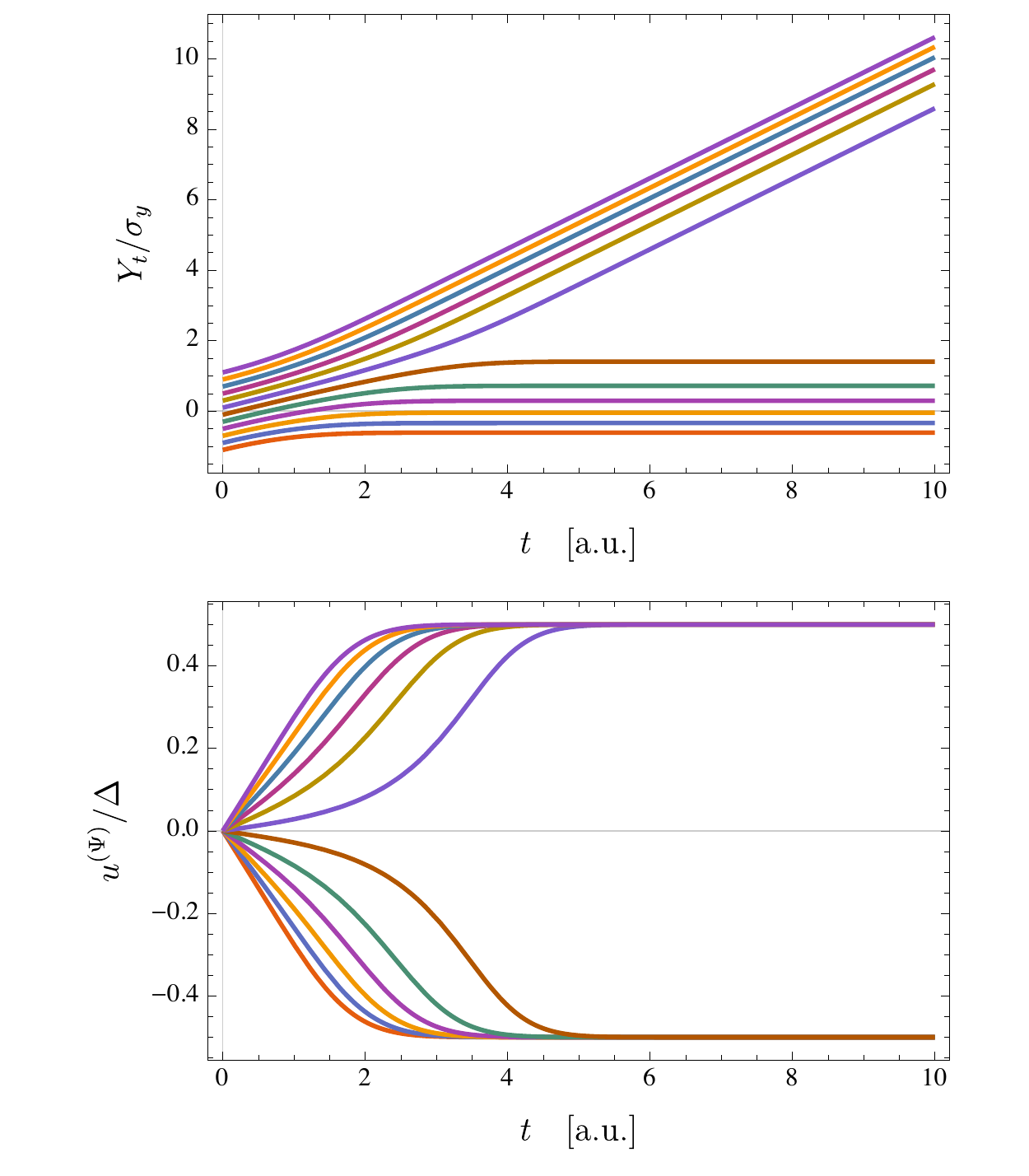}
	\caption{(a) Particle trajectories $ Y_t $  (normalized with the standard deviation) for evenly separated initial positions
		for the case of two entangled spin 1/2 particles. (b)
		The conditional energy change corresponding to the trajectories in (a) normalized with the splitting parameter $  \Delta = \hbar^2 k^2 / (2m) $.}
	\label{fig:steering}
\end{figure}

For large enough $ vt $, the unitary $ \hat{U}_Y $ separates the wave function into two wave packets with disjoint $ y $ support,
one centred at $ (x,y) = (0,0) $ and another at $ (x,y)=(0,vt) $. Thus, at the end only one of the terms in Eq. \eqref{eq:rd_psi} is relevant for the
conditional wave function since either $ g(Y_t) \approx 0 $ or $ g(Y_t - vt) \approx 0 $. During the rotation $ \hat{U}_Y $ no interaction term exists
between the particles and the Hamiltonian of particle $Y$ has no explicit time dependence. Thus, the only contribution comes from entanglement.
To get the trajectories we use the continuity equation. The probability distribution over the configuration space, $\mu(x,y,t) := |\Psi(x,y,t)|^2$, is given by summing the distributions associated with the individual spin components~\cite{teufel2009bohmian}. In this case, $\mu(x,y,t) = \sumss\Psi_{s,s'}^*(x,y,t)\Psi_{s,s'}(x,y,t) = g^2(y+vt)f^2(x) / 2 + g^2(y)f^2(x) / 2$, where $\Psi_{s,s'}(x,y,t)$ is the $\ket{s_X}\ket{s'_Y}$ component of $\Psi$. Thus,
\begin{align}
	\partial_t \mu(x,y,t)
	 & = -v g(y+vt)[ \partial_y g^2(y+vt) ] f^2(x) \nonumber                                                \\
	 & = \partial_y [ v \frac12g^2(y+vt)f^2(x) ] \nonumber                                                  \\
	 & = - \partial_y \left(\frac{v \mu_{\uparrow\uparrow}(x,y,t)}{\mu(x,y,t)} \mu(x,y,t) \right) \nonumber \\
	 & = - \partial_y [ v_y(y,t) \mu(x,y,t) ],
\end{align}
where $\mu_{\uparrow\uparrow}(x,y,t) = g^2(y+vt)f^2(x) / 2$ and
$v_y(y,t) = v g^2(y + vt)/ [g^2(y + vt) + g^2(y)] $. With the condition that the probability current
density vanishes at infinity, this implies that the velocity field for the $ Y $ particle is given by $ v_y(y,t) $.
Typical trajectories for the $Y$ particle are plotted in Fig. \ref{fig:steering}(a) for different initial conditions.

The conditional energy is evaluated for the kinetic energy of particle $X$, since no other potential is present, which yields
\begin{align}
	 & u\up(t|\vec{z}_0) \nonumber                                                                                                              \\
	 & = -\frac{\hbar^2}{2m}\frac{\int\dd x\ \sumss\Psi^*_{s,s'}(x,Y_t,t) \partial_x^2\Psi_{s,s'}(x,Y_t,t))}{\int\dd x\ \mu(x,Y_t,t)} \nonumber \\
	 & = -\left[\frac{\hbar^2 g^2(Y_t-vt)}{4m}\int\dd x e^{\imath k x}f(x) \partial_x^2(e^{-\imath k x}f(x)) \right. \nonumber                  \\
	 & \left.\quad + \frac{\hbar^2 g^2(Y_t)}{4m}\int\dd x f(x) \partial_x^2f(x) \right] \frac{1}{\int\dd x \mu(x,Y_t,t)} \nonumber              \\
	 & = \frac{\hbar^2}{2m} \left[ \frac{1}{4\sigma_X^2} + k^2 \frac{g^2(Y_t -vt)}{g^2(Y_t -vt) + g^2(Y_t)}\right],
\end{align}
where $\hat{P}_X$ is the momentum operator and $m$ the mass of particle $X$.

The main results here can be easily understood on physical grounds. Initially, the energy is independent of $ y $ and simply the
average of the energy of a
Gaussian wave packet with zero group velocity, $ E^{(0)} = \hbar^2 / 8 m \sigma_X^2 $ and that of a Gaussian wave packet with
group velocity $ k $, $ E^{(k)} = E^{(0)} + \hbar^2 k^2 / (2m) $.
As time evolves and the wave packets begin to separate, the energy changes depending on the exact trajectory $ Y_t $. These trajectories and the corresponding energies  are plotted in Fig. \ref{fig:steering}. Once the wave packets are well separated, only one term in Eq.~\eqref{eq:rd_psi} is relevant and the wave function behaves as if \textit{effectively factorized} (see Ref.~\cite{teufel2009bohmian} for a discussion on effective wave functions). As a consequence, the energy converges either to $ E^{(0)} $ or $ E^{(k)} $.  The fact that the spreading energy
$ E^{(0)} $ remains constant is a consequence of the assumption that we can neglect the spreading of the wave packets for the duration of $ \hat{U}_Y $. We also note that
the average energy remains constant at all times as expected since there is no interaction term.

\section{Conclusions and Discussion} \label{sec:conclusions}

In this paper we have determined the contributions to a quantum system's energy exchanges when it is coupled to an environment and externally driven. Going beyond the reduced system state picture, which cannot distinguish energy flows that arise from existing entanglement between system and environment or from their ongoing interactions, we have here used conditional wave functions that allow a single-shot analysis. Based on the CWF we have here derived a formally exact analytic expression for the energy exchanges of the system in a single run of an experiment, stated in Eq.~\eqref{eq:du}, without restricting how the global Hamiltonian is dependent on time or the form of the interaction Hamiltonian.

The derivation reveals three distinctly different contributions: an external contribution,
an interaction contribution, and an entanglement contribution,
directly associated with entanglement between the system and the environment.
Each of these contributions can be present on its own, e.g. when the system and environment are
entangled but not interacting and $\HS$ has no explicit time dependence, only the entanglement contribution is present.  Naturally, in order to entangle the system with its environment there must have been an interaction. However, after such initial preparation, the interaction Hamiltonian can be switched off and the interaction contribution vanishes, but the entanglement contribution remains. This provides a direct link between entanglement and energy fluctuations in a single run of an experiment for the first time.

Taking the statistical average for these contributions, Eqs.~\eqref{eq:du_int} - \eqref{eq:du_exp}, i.e. the average over many runs of the experiment, we have related the single-shot analysis to the expectation value of $\HS(t)$, Eq.~\eqref{eq:ev-evv}, and its time derivative,
Eqs.~\eqref{eq:exener} - \eqref{eq:evvdotent}. The external contribution yields the familiar expectation value of the Hamiltonian's explicit time dependence.
The term containing the time dependence of the reduced density operator splits into the average of the interaction and entanglement contributions, Eq.~\eqref{eq:entener}, where the average of the entanglement contribution can only be nonzero if an interaction is present, in contrast to the single-shot case. This is in line with the expectation that there can be no average energy transfer due to entanglement alone. We have demonstrated the results with a number of concrete examples that help to provide an intuitive picture of energy flow in physical space.

CWFs are closely related to weak values~\cite{AAV88,Flack_2018,norsen14} and can be reconstructed experimentally~\cite{Mahler2016, Kocsis2011,Xiao2017}, making the conditional energy empirically accessible.
An interesting open question is if the entanglement contribution could be experimentally used to quantify ``quantumness'' i.e. the appearance of entanglement between the system and the environment, and to what extent it may be linked to quantum advantages in thermodynamic processes.

Generalizing the presented results for the bare Hamiltonian to investigate the fluctuations associated with the interaction term in the Hamiltonian could open new methods to tackle strongly coupled quantum systems, and connect with known thermodynamic results. For example, the CWFs would allow one to identify effective energetic exchanges when one considers coarse-graining methods and effective Hamiltonians such as the Hamiltonian of mean force~\cite{kirkwood,Campisi2009}. Given the success in addressing quantum transport in nanoelectronic systems, it would be interesting to relate the statistical description given here to thermodynamic notions of work and heat in these systems~\cite{Ludovico16}. Furthermore, the quantum to classical transition could also be studied by generalizing classical results~\cite{Philipp17,Miller17} and analysing the limit of large quantum numbers, vanishing entanglement or centre-of-mass dynamics~\cite{Benseny17}.

\section{Acknowledgments}
This work has been supported in part by the Academy of Finland through its Quantum Technology Finland CoE project No. 312298. R.S. acknowledges the financial support by the Magnus Ehrnrooth Foundation as well as from the CMMP Education Network. J.A. acknowledges support from EPSRC (grant EP/M009165/1) and the Royal Society. This research was supported by the COST network MP1209 ``Thermodynamics in the quantum regime".

\bibliography{qthermo}

\appendix

\section{Coarse-graining for mixed states}\label{sec:app}

Following Ref.~\cite{Durr05}, Bohmian-like trajectories can equally be defined for unitary evolution of general mixed states
$ \hat{\sigma}(t) $. Indeed, for Hamiltonians of the form $ \hat{H} = \sum_{i=0}^N\sum_{j=1}^3 ({\hat{P}_j^{(i)^2}}/2m^{(i)} ) + V(\{\hat{Z}\ui_j\}, t) $ the
velocity field is formally equal to Eq.~\eqref{eq:vi_cwf} in the main text, for $ \hat\sigma(t) $ now a general mixed state rather than a pure state.
The trajectories are then defined formally in the same way as in the main text.
The natural generalization of the conditional energy is taken from Eq.~\eqref{eq:condenergy} in the main text, namely
\begin{equation}\label{eq:condenergy_mixed}
	u^{(\sigma)}(t|\vec{z}_0) = \frac{\Tr{\HS(t) {\hat{\tilde{\rho}}}^{(\sigma)}(t|\vec{z}_0)}}{\Tr{\hat{\tilde{\rho}}^{(\sigma)}(t|\vec{z}_0)}},
\end{equation}
where $ \hat{\tilde{\rho}}\us(t|\vec{z}_0) := \evys{\hat{\sigma}(t)} $ defines an unnormalized conditional density matrix, the
generalization of the conditional wave function \cite{norsen14}, and $ \vec{Y}\us_t := \vec{Y}\us(\vec{z}_0, t) $ is a trajectory of the
environment generated by $ \hat\sigma(t) $ given the initial condition $ \vec{z}_0 $ using a notation analogous to the main text.
The derivation now follows \textit{verbatim} and we find the generalized quantities
\begin{align}
	\dd_t u\us(t|\vec{z}_0) = \dd_t{u}\us_{\text{int}}(t|\vec{z}_0)
	+ \dd_t{u}\us_{\text{ent}}(t|\vec{z}_0)
	+ \dd_t{u}\us_{\text{ext}}(t|\vec{z}_0),
\end{align}
where
\begin{align}
	\dd_t{u}\us_{\text{int}}(t|\vec{z}_0) = & \frac{\Tr{\dd_t{\hat{\tilde\rho}}\us_{\text{int}}(t|\vec{z}_0)[\HS(t) - u\us(t|\vec{z}_0)]}}{\Tr{\hat{\tilde\rho}\us(t|\vec{z}_0)}}; \label{eq:du_int_mixed} \\
	\dd_t{u}\us_{\text{cor}}(t|\vec{z}_0) = & \frac{\Tr{\dd_t{\hat{\tilde\rho}}\us_{\text{cor}}(t|\vec{z}_0)[\HS(t) - u\us(t|\vec{z}_0)]}}{\Tr{\hat{\tilde\rho}(t|Y_t)}}; \label{eq:du_ent_mixed}          \\
	\dd_t{u}\us_\text{ext}(t|\vec{z}_0) =   & \frac{\Tr{\hat{\tilde\rho}\us(t|\vec{z}_0) \dd_t{\hat{H}}_{\rm S}(t)}}{\Tr{\hat{\tilde\rho}\us(t|\vec{z}_0)}},
\end{align}
with $ \dd_t{\hat{\tilde{\rho}}}\us_{\text{int}}
	= -\frac{\imath}{\hbar}  \evys{[\HI, \hat{\sigma}(t)]}$ and $  \dd_t{\hat{\tilde{\rho}}}\us_\text{cor} =  -\frac{\imath}{\hbar}
	( \evys{[\HE, \hat{\sigma}(t)]} - v^{(y)} \cdot \evys{[\hat{\vec{P}}^{(y)}, \hat{\sigma}(t)]}) $.
The only difference is that the entanglement term should now be understood as a correlations term since it will be
nonzero for both classical and quantum correlations.
Finally, the relations for the statistical averages also hold, namely
\begin{equation}
	\evv*{u\us(t)} := \int\dd\mathbf{z} \ev*{\hat{\sigma}(0)}{\vec{z}} u\us(t|\vec{z}) =
	\Tr{\HS(t)\hat{\sigma}(t)},
\end{equation}
and
\begin{align}
	\Tr{\hat{\sigma}(t)\dd_t\HS(t)}  & = \evv*{\dd_t{u}\us_\text{ext}(t)};                                                                   \\
	\Tr{\HS(t)\dd_t\hat{\sigma}(t)}  & =
	\evv*{\dd_t{u}\us_\text{int}(t)}
	+ \evv*{\dd_t{u}\us_\text{ent}(t)};                                                                                                      \\
	\evv*{\dd_t{u}\us_\text{ent}(t)} & = \int\dd \mathbf{z}\ u\us(t,\vec{y})\ \ev*{\dd_t{\hat{\tilde\rho}}\us_\text{int}}{\vec{x}} \nonumber \\
	                                 & = -\frac{\imath}{\hbar}\Tr{\hat{\sigma}(t) [\hat{u}\us(t), \HI]}.
\end{align}
\end{document}